\documentclass[11pt]{article}

\usepackage{cite}
\usepackage{amssymb}
\usepackage{amsmath}
\usepackage{bm}
\usepackage{graphicx}
\usepackage{hyperref}
\begin{document}

\title{Dirac, Bohm and the Algebraic Approach}
\author{B. J. Hiley\footnote{E-mail address b.hiley@ucl.ac.uk.} and G. Dennis}
\date{Physics Department, University College, Gower Street, London
WC1E 6BT.\\TPRU, Birkbeck, University of London, Malet Street, \\London WC1E 7HX.\\ \vspace{0.4cm} }
\maketitle

\begin{abstract}
In this paper we show how Dirac, in 1947, anticipated the Bohm approach using an  argument based on what is now called the Heisenberg picture.  From a detailed examination of these ideas, we show that the role played by the Dirac standard ket is equivalent to the introduction and use of the idempotent in the orthogonal and symplectic Clifford algebras.  This formalism is then used to show that the so-called `Bohm trajectories' are the average of an ensemble of individual stochastic Feynman paths.  Since the Bohm approach can be simply reduced to classical mechanics, the algebraic formalism presented here provides a natural way to relate quantum mechanics to classical mechanics without the need  for decoherence.  We show that this approach suggests an underlying fractal space-time of the type discussed by Nottale.

\end{abstract}

\section{Introduction}

Dirac~\cite{pd65} has argued that in quantum field theory with its infinite degrees of freedom  ``the Heisenberg picture is a good picture, the Schr\"{o}dinger picture is a bad picture, and the two pictures are not equivalent".  The equivalence has only been established in particle theory where the number of degrees of freedom is finite.  But even this equivalence is a mathematical one, established up to  a unitary transformation.  To generate the Heisenberg picture, we use the unitary transform  $U(t)=\exp[iHt/\hbar]$, where $H$ is the Hamiltonian of the system.  However it is not clear that mathematical equivalence necessarily gives a physically equivalent picture. Therefore a further more detailed examination of what has been called `matrix mechanics' should be carried out to clarify the physical implications of the different pictures.   

By `matrix mechanics' we mean that the dynamics is carried by an algebra of operators or $q$-numbers as Dirac calls them. These
 elements are time dependent and all the dynamics is therefore contained in the $q$-numbers themselves.  Here  the wave function or ket plays a passive role, being independent of time and fixed at some initial time, $t_0$, and so it can be written as $| \psi(t_0)\rangle$.  Indeed in the Fock representation, this ket becomes the vacuum state defined by $a|0\rangle=0$.  If we choose $| \psi(t_0)\rangle$ to be the state of zero energy, it can be regarded as playing the role of a vacuum state so that $\hat P|\psi(t_0)\rangle=0$. Moreover Schwinger~\cite{js53} has pointed out that when using the
 Heisenberg picture in quantum field theory [QFT], one must always define the vacuum state because in QFT  there can exist many inequivalent vacuum  states.
This means that vacuum states have a key role to play~\cite{bh15}.

Dirac~\cite{pd47}, in his classic text, has  already provided a symbolic
 means of distinguishing vacuum states when he introduces a new stand alone symbol, $\rangle$, which he calls the {\em standard ket}.    This is not to be confused with  the more familiar form of ket $|\psi\rangle$. Notice there is a mathematical distinction between these two objects.  Removal of the $|$ enables us to work completely within the $q$-algebra.  The symbol $\;\rangle$ prevents multiplication from the right and so  Dirac has  actually constructed an element of a left ideal in the algebra.  It is this element that carries all the information normally carried by the wave function.  An earlier discussion of this symbol in an algebraic context can be found in Frescura and Hiley~\cite{ffbh84}.
The wave function as a vector in an added Hilbert space is therefore unnecessary but one can recover the usual representation in such a space if needed. However we should point out that Dirac~\cite{pd65} argued that something more general is needed as in QED it is not possible to represent the dynamical variables as matrices or as operators in a Hilbert space.  We will not discuss this generalisation in this paper.
  
How then do we move from the algebraic standard ket to the usual ket?  Let us recall Dirac's argument by first defining the relation between the standard ket $\rangle$ and the conventional ket $|\;\;\rangle$.  To do this Dirac introduces a new form of delta function defined by 
\begin{eqnarray*}
\delta_{\hat A,a_i }\rangle=|a_i\rangle\hspace{1cm}\mbox{or}\hspace{1cm}\delta(\hat A-a_i)\rangle=|a_i\rangle
\end{eqnarray*}
depending on whether the set of  eigenvalues, $a_i$, of the $q$-number $\hat A$ are discrete or continuous\footnote{$q$-numbers satisfy two binary requirements, addition and multiplication, the latter being non-commutative.}.  Then we can write, 
\begin{eqnarray*}
\sum_{a_j}\langle a_j|\hat A\delta_{\hat A,a_i}\rangle=a_i\hspace{1cm}\mbox{or}\hspace{1cm}\int\langle a_j|\hat A\delta(\hat A-a_i)\rangle da_j=a_i.
\end{eqnarray*}
With any function  of $q$-numbers, $\psi({\hat A})$,  we can associate the element $\psi(\hat A)\delta_{\hat A,a_i}\rangle$ or $ \psi(\hat A)\delta(\hat A-a_i)\rangle$ so that
\begin{eqnarray}
\sum_{a_j}\langle a_j|\psi(\hat A)\delta_{\hat A,a_i}\rangle=\psi(a_i)     \hspace{0.5cm}\mbox{or}\hspace{0.5cm}
\int \langle a_j|\psi(\hat A)\delta(\hat A-a_i)\rangle da_j=\psi(a_i).	\label{eq:bra-ket}
\end{eqnarray}
Thus in more general terms we can write
\begin{eqnarray*}
\delta_{\hat A_1,a_1}\delta_{\hat A_2,a_2}\dots \delta(\hat A_r-a_r)\delta(a\hat A_{r+1}-a_{r+1})\dots\rangle=|a_1,a_2\dots a_r,a_{r+1}, \dots\rangle.
\end{eqnarray*}
This generalisation shows how a collection of $q$-numbers, some having a discrete basis while the rest have a continuous basis, is related to the usual ket.  

  As we have remarked the symbol $\rangle$ was introduced to prevent multiplication from the right thereby forming a left ideal in the algebra.  
In order to allow right multiplication, Dirac introduces an algebraic dual to the standard ket, namely the standard bra, $\langle\;$, and defines its relation to the conventional bra by
\begin{eqnarray*}
\langle \delta_{\hat A_1a_1}\delta_{\hat A_2a_2}\dots \delta(\hat A_r-a_r)\delta(\hat A_{r+1}-a_{r+1})\dots =\langle a_1, a_2\dots a_r, a_{r+1}\dots|.
\end{eqnarray*}
The complex conjugate of the wave function is therefore contained in a right ideal of the algebra.  Technically this means that we have constructed a two sided, left/right module. 

 In equation~(\ref{eq:bra-ket}) we have constructed a symbol $\langle\;\;\rangle$ which is assumed to be a complex number so that  $\langle a|\psi\rangle=\psi(a)$ which is a conventional wave function.   More generally $\langle\phi|\psi\rangle$ is regarded as a transition amplitude [TA] from which we calculate the probability of a transition by forming $|\langle\phi|\psi\rangle|^2$.  In this sense, the wave function itself is a TA, $\psi\rightarrow\psi$, and not a function of state as usually assumed.  Seen in this light the quantum algebra becomes a way to describe fundamental processes as Feynman~\cite{rf:process} originally proposed.

 If we write $\rangle\langle$ , the two symbols taken together produce an idempotent, $\epsilon$, provided we define  $\langle\;1\;\rangle=1$. Then we have
\begin{eqnarray*}
\rangle\langle\;\;\rangle\langle\;\;=\;\;\rangle\langle 1 \rangle \langle\;\;=\;\;\rangle\langle\;.
\end{eqnarray*}
Why do we need an idempotent?  As we have already remarked above, in a Fock representation  it is not sufficient to introduce a set of creation and annihilation operators; we also need to add the projector, $V$, onto the vacuum with $V^2=V$.  In the context of a process algebra, $V$ becomes an idempotent element of the algebra.  

Dirac has the Heisenberg algebra, $(\hat X, \hat P)$, in mind when he adds the new symbol $\;\rangle$ with which we form an idempotent.  Thus writing $\epsilon=\;\rangle\langle\;$, we have $\epsilon^2=\epsilon$.
Since the Fock representation is related to the  Schr\"{o}dinger representation by the Segal-Bargmann transformation, this idempotent is clearly related to the vacuum projector $V=|0\rangle\langle0|$~\cite{ffbh84}.

In his later lectures, Dirac~\cite{pd12} changed the symbol for the
 standard ket and writes $|S\rangle$. Furthermore he clarifies the role of these symbols by distinguishing between two standard kets $|S_q\rangle$ and $|S_p\rangle$ in the Schr\"{o}dinger picture.  In this way we can distinguish between the position representation and the momentum representation in the Heisenberg picture.  The position representation is given by
\begin{eqnarray*}
|\psi\rangle =\psi(\hat Q)\;|S_q\rangle,\quad \hat P|S_q\rangle=0.
\end{eqnarray*}
While the momentum representation is given by
\begin{eqnarray*}
|\phi\rangle =\phi(\hat P)\;|S_p\rangle,\quad \hat Q|S_p\rangle=0.
\end{eqnarray*}
For the sake of completeness and to bring out the similarity we write the Fock representation in the form
\begin{eqnarray*}
|\psi\rangle=\psi(a^\dag)|0\rangle,\quad a|0\rangle=0.
\end{eqnarray*}
Incidentally we could also have 
\begin{eqnarray*}
|\phi\rangle=\phi(a)|P\rangle,\quad a^\dag|P\rangle=0.
\end{eqnarray*}
Here $| P \rangle$ is the `full state', namely the plenum.  Dirac~\cite{pd58}  introduces a similar idea which he denotes by the symbol $|0^*\rangle$.  For an interesting discussion on this structure see Finkelstein~\cite{df96}.

From the algebraic perspective we see that the Heisenberg Lie algebra defined by the pair $(\hat Q, \hat P)$ satisfying the commutator bracket $[\hat Q, \hat P]=i\hbar$ needs to be supplemented by adding an idempotent $|S\rangle\langle S|$.  The idempotent must be added from outside the algebra as the Heisenberg Lie algebra itself is nilpotent and therefore contains no idempotent. Indeed the algebra we have constructed is a symplectic Clifford algebra~\cite{ac90, ms57}. For a further discussion of this point see Hiley~\cite{bh01}.

This algebra is to be contrasted with the orthogonal Clifford algebra which must be used to extend the quantum formalism to the relativistic domain so as to include spin.  Hiley and Callaghan~\cite{bhrc10a, bhrc10b} have shown how  the 
Pauli and Dirac Clifford algebras can be used to extend the Bohm approach to the relativistic domain.  The orthogonal Clifford algebras are non-nilpotent and contain many idempotents. These idempotents are unitarily  equivalent and emphasise different physical aspects of the process.  For example in the Pauli algebra, the idempotent $e=(1+\sigma_x)/2$ will pick out the $x$-direction if this direction is defined by an externally applied homogeneous magnetic field, the field breaking the rotational symmetry.

In his lecture notes, Dirac~\cite{pd12} introduces a standard ket to play the same role as we will now demonstrate.  The standard ket $|S_1\rangle$ in this case is defined by the relation
\begin{eqnarray*}
\sigma_1|S_1\rangle=|S_1\rangle.
\end{eqnarray*}
Dirac then shows that in the representation in which the Pauli matrix $\sigma_3$ is diagonal
\begin{eqnarray*}
|S_1\rangle =   \frac{1}{\sqrt2} \begin{pmatrix} 
      1 \\
      1 \\
   \end{pmatrix}.
\end{eqnarray*}
We can then form
\begin{eqnarray*}
|S_1\rangle\langle S_1|=  \frac{1}{2} \begin{pmatrix} 
      1 & 1 \\
      1 & 1 \\
   \end{pmatrix}=(1+\sigma_1)/2
\end{eqnarray*}
which is just the idempotent used to break the symmetry in the $x$-direction.  

In this paper we will show that the introduction of the idempotent into the Heisenberg algebra produces a very different way of looking at the quantum formalism - a way that was already implicit in Dirac's classic works~\cite{pd27, pd47}.  However the real surprise is that we can thereby  make contact with the Bohm approach~\cite{dbbh93}, a point which we will bring out later in the paper.

Before leaving this topic we would like to draw attention to a series of papers by Sch\"{o}nberg~\cite{ms58} on the topic of ``Quantum Mechanics and Geometry" which inspired one of us [BJH] to explore an algebraic approach in the first place.  The motivation was two-fold.  Firstly in the sixties when BJH joined him, Bohm was exploring the possibility of understanding quantum phenomena in terms of what he called `structure process'~\cite{db65}.  At the same time Penrose was developing his ideas on spin networks~\cite{rp71} and twistor theory~\cite{rp67} with an aim of generating a quantum geometry.
A detailed discussion of this mathematical structure can be found in the papers of 
 Frescura and Hiley~\cite{ffbh84} and Hiley~\cite{bh14, bh11}.



\section{The Dirac-Bohm Picture}

Let us recall how the addition of a single idempotent to the Heisenberg algebra works in practice.   Any element of a left ideal can be written as $\psi(\hat X)\epsilon$ so that
\begin{eqnarray*}
\psi(x)=\int\langle x'|\psi(\hat X)\delta(\hat X-x)\rangle dx'.
\end{eqnarray*}
But what happens when an element of the quantum algebra, $\hat A$, is a differential such as $d/d\hat X$? We can then simply write
\begin{eqnarray*}
\frac{d}{d\hat X}\psi(\hat X)|S_x\rangle=\frac{d\psi(\hat X)}{d\hat X}|S_x\rangle
\end{eqnarray*}
where $|S_x\rangle$ is the standard ket in the $x$-representation.
Since this holds for all functions $\psi(\hat X)$, including $\psi(\hat 1)=1$, we must have
\begin{eqnarray*}
\frac{d}{d\hat X}|S_x\rangle=0.
\end{eqnarray*}
This begins to bring out the similarity with the vacuum state $|0\rangle$.  Here $d/d\hat X$ is behaving like an annihilation operator, $a|0\rangle=0$. Whereas $\hat X$ behaves like a creation operator, $a^\dag|0\rangle=|1\rangle$.

 To further support this similarity,  we follow Dirac~\cite{pd47} and consider what happens when we apply a derivative to a standard bra, $\langle S_x|{\frac{\overleftarrow d}{d\hat X}}$. For consistency we must have
 \begin{eqnarray*}
\left\{\langle S_x|\phi(\hat X)\frac{ \overleftarrow d}{d\hat X}\right\}\psi(\hat X)|S_x\rangle=\langle S_x|\phi(\hat X)\left\{\frac {\overrightarrow d}{d\hat X}\psi(\hat X)|S_x\rangle\right\}
\end{eqnarray*}
so that we can now write this as
\begin{eqnarray*}
\int_{-\infty}^\infty\langle S_x|\phi(\hat X)\frac{\overrightarrow d}{d\hat X}|x'\rangle dx'\psi{(x')}=\int_{-\infty}^\infty\phi(x')dx'\frac{d\psi(x')}{dx'}.
\end{eqnarray*}
Let us now perform partial integration on the right-hand side of this equation to obtain
\begin{eqnarray*}
\int_{-\infty}^\infty\langle S_x|\phi(\hat X)\frac{\overrightarrow d}{d\hat X}|x'\rangle dx'\psi{(x')}=-\int_{-\infty}^\infty\frac{d\phi(x')}{dx'}dx'\psi(x')
\end{eqnarray*}
provided that the contribution from the limits vanishes.  This gives
\begin{eqnarray*}
\langle S_x|\phi(\hat X)\frac{d}{d\hat X}|x'\rangle =-\frac{d\phi(x')}{dx'},
\end{eqnarray*}
showing that
\begin{eqnarray*}
\langle S_x|\phi(\hat X)\frac{\overleftarrow d}{d\hat X}=-\langle S_x|\frac{d\phi(\hat X)}{d\hat X}.
\end{eqnarray*}

Now Dirac~\cite{pd47} observes that the complex conjugate of the linear operator $d/d\hat X$ can be found by noting that the  conjugate imaginary of $d\psi(\hat X)/d\hat X|S_x\rangle$ can be written in the form $-\langle S_x| \bar\psi(\hat X)\overleftarrow d/d\hat X$.  This shows that the complex conjugate of $\overleftarrow d/d\hat X$ is $-\overrightarrow d/d\hat X$.  We introduce the over-bar arrows to emphasise that our mathematical structure is a bimodual. 

We can now work out the  commutation relations between $d/d\hat X$ and $\hat X$. Thus
\begin{eqnarray*}
\frac{d}{d\hat X}(\hat X\psi(\hat X))|S_x\rangle=\psi(\hat X)|S_x\rangle +\hat X\frac{d}{d\hat X}\psi(\hat X)|S_x\rangle.
\end{eqnarray*}
Since this holds for any  function  $\psi(\hat X)$, we have
\begin{eqnarray}
\frac{d}{d\hat X}\hat X-\hat X\frac{d}{d\hat X}=1.		\label{eq:PQCom}
\end{eqnarray}
If we write $\hat P=-i\hbar d/d\hat X$, we see that we are looking specifically at the Schr\"{o}dinger representation.

An important lesson that we learn from repeating the Dirac argument here  is that it is the non-commuting algebra of $q$-numbers that is the essential structure.  The resulting Hilbert space is merely a representation and, in our view,  restricting ourselves to only one specific representation, namely the Schr\"{o}dinger representation, provides only a partial understanding of quantum phenomena.    Thus any attempt to construct a `picture of reality' based solely on the Schr\"{o}dinger representation alone will not be rich enough to capture the deeper subtleties of quantum phenomena.

\section{Algebraic Elements in Polar Form}

\subsection{The Algebraic Schr\"{o}dinger Equation}

Dirac~\cite{pd47} shows  that the algebraic form of Schr\"{o}dinger's equation can be written as
\begin{eqnarray}
i\hbar \frac{\partial}{\partial t}\psi(\hat X,t)|S_x\rangle=H\psi(\hat X,t)|S_x\rangle.	\label{eq:SE}
\end{eqnarray}
Notice $\psi(\hat X, t)$ is a $q$-number acting on the standard ket $|S_x\rangle$.  Here the Hamiltonian is time independent.
Now we want to enquire how the elements, $\psi(\hat X,t)$, develop in time when they are written in polar form
\begin{eqnarray*}
\psi(\hat X,t)=R(\hat X,t)\exp[iS(\hat X,t)/\hbar],
\end{eqnarray*}
where $R$ and $S$ clearly commute and are real functions of the $q$-number $\hat X$  and  the parameter $t$.
Putting this expression into equation (\ref{eq:SE}), we find
\begin{eqnarray}
\left\{i\hbar\frac{\partial R}{\partial t}-R\frac{\partial S}{\partial t}\right\}|S_x\rangle=e^{-iS/\hbar}H(\hat X, \hat P)e^{iS/\hbar}R|S_x\rangle.	\label{eq:2}
\end{eqnarray}
Applying the unitary $q$-number $e^{iS/\hbar}$ to the Hamiltonian $H(\hat X, \hat P)$,  $\hat X$ is unchanged but 
\begin{eqnarray*}
e^{-iS/\hbar}\hat P e^{iS/\hbar}= \hat P +\frac{\partial S}{\partial \hat X}
\end{eqnarray*}
so that equation (\ref{eq:2}) becomes
\begin{eqnarray}
\left\{i\hbar\frac{\partial R}{\partial t}-R\frac{\partial S}{\partial t}\right\}|S_x\rangle=H\left(\hat X, \hat P+\frac{\partial S}{\partial \hat X}\right)R|S_x\rangle.	\label{eq:SE2}
\end{eqnarray}

To see precisely what happens to the right-hand side of this expression we will use an oscillator Hamiltonian of the form $\hat X^2+\hat P^2$. Thus 
\begin{eqnarray*}
H\left(\hat X, \hat P+\frac{\partial S}{\partial \hat X}\right)R|S_x\rangle=\left(\hat X^2R-\hbar^2\frac{\partial^2R}{\partial\hat X^2}+\left(\frac{\partial S}{\partial\hat X}\right)^2R-2i\hbar\frac{\partial S}{\partial \hat X}\frac{\partial R}{\partial \hat X}-i\hbar\frac{\partial^2S}{\partial\hat X^2}R\right)|S_x\rangle.
\end{eqnarray*}
If we separate the real and imaginary parts, we find the imaginary part gives
\begin{eqnarray*}
\left(R\frac{\partial R}{\partial t}+2\frac{\partial S}{\partial \hat X}\frac{\partial R}{\partial \hat X}+\frac{\partial^2S}{\partial\hat X^2}R\right)|S_x\rangle=0
\end{eqnarray*}
or
\begin{eqnarray}
\left(\frac{\partial \rho}{\partial t}+\frac{\partial}{\partial\hat X}\left(\rho\frac{\partial S}{\partial\hat X}\right)\right)|S_x\rangle=0.		\label{eq:CP1}
\end{eqnarray}
The real part is
\begin{eqnarray}
\left(\frac{\partial S}{\partial t}+\left(\frac{\partial S}{\partial\hat X}\right)^2-\hbar^2\left(\frac{\partial^2R}{\partial\hat X^2}\right)\frac{1}{R}+\hat X^2\right)|S_x\rangle=0.	\label{eq:QHJ1}
\end{eqnarray}
If we now multiply equation (\ref{eq:CP1}) by $\langle S_x|\delta(\hat X-x)$,   we find
\begin{eqnarray}
\frac{\partial \rho}{\partial t}+\frac{\partial }{\partial x}\left(\rho\frac{\partial S}{\partial x}\right)=0.		\label{eq:CP2}
\end{eqnarray}
Multiplying equation (\ref{eq:QHJ1}) by the same factor produces
\begin{eqnarray}
\frac{\partial S}{\partial t}+\left(\frac{\partial S}{\partial x}\right)^2-\frac{\hbar^2}{R}\left(\frac{\partial^2 R}{\partial x^2}\right)+x^2=0.  \label{eq:QHJ2}
\end{eqnarray}
Equation (\ref{eq:CP2}) will be recognised as the classical Liouville equation showing the conservation of probability if we identify the momentum $p$ with $\partial S/\partial x$.  For the harmonic oscillator, the quantum probability conservation equation is exactly the same as the classical Liouville equation.  Furthermore Dirac~\cite{pd47} shows that this result is generally true if we consider the case of $\hbar\rightarrow 0$.

Dirac did not write down equation (\ref{eq:QHJ2}) but if he had he would have noticed an extra term, the quantum potential.  Instead Dirac simply let $\hbar\rightarrow 0$, obtaining the classical Hamilton-Jacobi equation 
\begin{eqnarray}
\frac{\partial S}{\partial t}+H_c
\left(x,\frac{\partial S}{\partial x}\right)=0.	\label{eq:CHJ}
\end{eqnarray}
Having arrived at the classical Hamilton-Jacobi equation, why not go on to examine what would happen if one retains all powers of $\hbar$?  It appears Dirac did but came to this conclusion~\cite{pd47}:
\begin{quote}
By a more accurate solution of the wave equation one can show that the accuracy with which the coordinates and momenta simultaneously have numerical values cannot remain permanently as favourable as the limit allowed by Heisenberg's principle of uncertainty\dots
\end{quote}
In 1952 Bohm~\cite{db52} made a polar decomposition of the {\em wave function} in the $x$-representation and using the standard Schr\"{o}dinger equation, found exactly the same equations that Dirac had found.  Bohm then explored what happened if one did retain all terms in $\hbar$.  He noticed that the canonical momentum, $p_B=\nabla S$, did not correspond to the measured value of momentum and so he attributed it to the actual {\em local} momentum of the particle.  The measured momentum is the momentum obtained by integrating the local momentum suitably weighted over all space.  We will use $p_B$ to distinguish the {\em local} momentum from the {\em measured} momentum $p$.   

 Traditionally the value of the momentum is determined from many measurements made on an ensemble of similarly prepared particles. Each measurement corresponds to one of the momentum eigenvalues, since each
measurement is a standard von Neumann one.  This leaves open the question as to what is the exact meaning of the Bohm local momentum $p_B$.  Surely you cannot have {\em two types} of momenta?  We will leave the meaning of these momenta as an open question which we will answer later.   In the meantime, we will follow Bohm and assume that $p_B$ is the momentum actually possessed by the particle, a beable as discussed in Bohm and Hiley~\cite{dbbh93}.

   We now know that $p_B$ is given by the real part of the  weak value of the momentum defined by the relation~\cite{rl05, hw07, bh12}
\begin{eqnarray*}
p_B(x)=Re\langle \hat P\rangle_w=Re\frac{\langle x|\hat P|\psi(x,t)}{\langle x|\psi(x,t)\rangle}.
\end{eqnarray*}
This weak value has been measured by Kocsis {\em et al.}~\cite{skbb11} for  photons and is being measured for atoms by Morley {\em et al.}~\cite{mjpe16}. 

Furthermore  it has been shown by Hiley~\cite{bh12} that 
the measured average value of the momentum is actually 
\begin{eqnarray*}
p=\int \rho(\bm r)p_B(\bm r) d^3\bm r.
\end{eqnarray*}
This point was first made by Mackey~\cite{gm63}.
Thus there is a difference between a local momentum and the global momentum which is an average over all space.  The difference between a local momentum and a global momentum has been discussed by Colosi and Rovelli~\cite{dccr09}.


\subsection{The Bohm Approach}

Bohm's original concern~\cite{db52} was that the standard approach to the quantum formalism gave no clear description of what could be going on between measurements.  In today's terms, we would say that there was no underlying ontology.  Bohm was not looking for an alternative to the usual formalism, but rather a way to look at the same formalism that would give us an ontology. 

Rather than look for an ontology, the major concern was to develop a new description of the quantum dynamics that would unify it with electromagnetism and, eventually, gravity.  Let us first recall the original programme set out by Schwinger~\cite{js51, js53} in this regard.
Quantum mechanics involves two distinct sets of hypotheses:- (i) linear operators and state vectors with the probability interpretation; and (ii) a non-commutative algebra with trace functions providing the probabilities.  It was Schwinger's aim to unite these two approaches with quantum dynamical laws that would find their proper expression in terms of what he called {\em transformation functions}, and what Feynman~\cite{rf48}  called {\em transition amplitudes}.  As we now know this led to the Feynman propagator or Green's function approach.

Bohm's initial approach~\cite{db52} was more limited and was proposed in a preliminary form to show that an alternative {\em interpretation} of the quantum formalism was possible.  This was contrary to the general consensus at the time that there was no possible alternative to the Copenhagen interpretation.

Bohm simply starts with the standard Schr\"{o}dinger equation and separates it into its real and imaginary parts under the polar decomposition of the wave function $\psi(x,t)=R(x,t)e^{iS(x,t)/\hbar}$.  The imaginary part then gives
\begin{eqnarray*}
\frac{\partial \rho}{\partial t} +\frac{\partial}{\partial x}\left(\rho\frac{\partial S}{\partial x}\right)=0
\end{eqnarray*}
which is identical to equation (\ref{eq:CP2}).  Note that no approximation to $\hbar$  is required. 
The real part of the decomposition becomes
\begin{eqnarray}
\frac{\partial S}{\partial t} +\frac{1}{2m}\left(\frac{\partial S}{\partial x}\right)^2-\frac{\hbar^2}{2mR}\left(\frac{\partial ^2R}{\partial x^2}\right)+V(x,t)=0.		\label{eq:QHJ}
\end{eqnarray}
Notice that equation (\ref{eq:QHJ})  reduces to equation (\ref{eq:QHJ2}) for an harmonic oscillator, indicating that the Schr\"{o}dinger equation already contains a dynamics that is much closer to that of the classical domain than at first imagined.

Of course Bohm knew that by neglecting the $\hbar^2$ term one obtains the classical Hamilton-Jacobi equation.  Indeed it was while exploring the WKB approximation that Bohm noticed  that, by keeping terms of all orders $\hbar$,  one can still use classical concepts such as the position and momentum of a particle and therefore retain the notion of a particle trajectory.  How, just by including  higher order terms in $\hbar$, could the conceptual structure change as dramatically as  the Copenhagen interpretation envisaged?  Is one really up against the uncertainty principle as stated by Dirac?

Recall the uncertainty principle states that it is not possible to {\em measure} the  position and momentum simultaneously.  Therefore one has  three possibilities.  (i) The quantum particle does not possess simultaneously a position and momentum.  (ii) The quantum particle does have a simultaneous position and momentum but we cannot say anything about them.  (iii) The quantum particle does have a simultaneous position and momentum, $(x,p)$, but it is just not possible to measure them simultaneously.
Adopting (iii) we can use $(x,p)$ as dynamical variables, the behaviour of which are governed by the $(R,S)$ appearing in the real and imaginary parts of the Schr\"{o}dinger equation. This was the position that Bohm and Hiley adopted in their book ``The Undivided Universe"~\cite{dbbh93}.

Earlier Philippidis, Dewdney and Hiley~\cite{cpcdbh79} assumed
 that the local momentum is given by $p_B=\nabla S$, and showed that by integrating this
expression using $S$ as obtained from the solution of the Schr\"{o}dinger equation, they were able to find sets of orbits which were tentatively identified with `particle trajectories'.  The technique has been applied to many quantum phenomena such as the two-slit experiment, quantum tunnelling etc., details of which will be found in~\cite{dbbh93}.  This gave a very different interpretation of quantum phenomena.
However there has always been a puzzle as to how it is possible to obtain a continuous `trajectory' without violating the uncertainty principle. Surely the underlying structure is itself non-commutative, so what do these orbits represent physically?  In the next sections we will provide a tentative answer to this question.  

Before providing the answer, we should point out that in this approach, measurement becomes a participatory process since a measurement of one variable changes all the complementary variables. Thus if the particle is in a state $\psi(\bm r,t)$, then a measurement of momentum $p$ gives an average value, $p=\int \rho(\bm r, t)p_B(\bm r,t) d^3\bm r$, where the integral is taken over all space.  This agrees with the result of standard quantum mechanics so there is no contradiction. A detailed explanation for how this works will be found in the chapter on measurement in~\cite{dbbh93}.

\subsection{How is the Dynamical Evolution Affected by Non-commutativity?}

Now it is necessary to find the actual meaning of the orbits discussed in the previous section in the context of a non-commutative underlying structure.  This question was first raised by Dirac~\cite{pd45} where he remarks that the close analogy with classical mechanics could be seen simply by making the variables of classical mechanics into non-commuting $q$-numbers.  At that time he was hampered by the fact that there were very few mathematical techniques involving non-commuting variables available. 

It is clear that the non-commutativity arrived at by Dirac through equation (\ref{eq:PQCom}) is very basic since it involves position and displacement, which apply to all movement, classical and quantum.  Nothing specifically quantum has so far been invoked. However just as the particle is obtained from the vacuum, we can regard  position as similarly being obtained, not from the vacuum, but from $\epsilon$, an object unitarily related to the vacuum under the Segal-Bargmann transformation. (See Frescura and Hiley~\cite{ffbh84}.)

Adopting the spirit of the Frescura-Hiley approach, we can argue that position and momentum can be constructed from the `vacuum' $\epsilon=\;\rangle\langle\;.$
Thus the position is obtained from the expression $x=\int\langle x'| \hat X\delta(\hat X-x)\;\rangle dx'$.  The coordinate is thereby constructed from $\epsilon$ itself.  Our algebra is not in space, but has within it the potential for {\em constructing} a space.  However we know that there may be many equivalent idempotents in an algebra, and so there is the potential  for creating many spaces, only one of which is realised at a time.  Hence the algebraic order, or the implicate order as Bohm~\cite{db80} preferred to
 call it, contains within it many explicate orders -- each order corresponding to a particular primitive idempotent, a particular vacuum state, corresponding to a given experimental arrangement.  
 
 Following Schwinger~\cite{js53} we think of the vacuum state as being an energy ground state and as remarked earlier, in quantum field theory we have the possibility of many inequivalent vacuum states.  The spontaneously broken vacuum state that is associated with the Higgs mechanism~\cite{ph66} is just one such example.  Ground states are familiar in solid
  state physics and indeed the notion of  symmetry breaking can be traced back to the work of Anderson~\cite{pa63}.

At this stage we should proceed cautiously and try to understand how the local momentum arises within this non-commutative geometry.  For this we return to Dirac and recall how the derivative $d/dx$ emerges from $\epsilon$.  This means that geometry emerges from $\epsilon$ itself.  In this way the algebra contains implicitly within it a whole ensemble of explicate geometries.  This then forms the basis of a quantum space-time geometry.

To bring in the momentum, Dirac introduces a transition amplitude [TA]
\begin{eqnarray*}
\langle x'_t|x''_{t_0}\rangle=e^{iS/\hbar},
\end{eqnarray*}
where $S$ is the classical action.  At this stage the reason for introducing the classical action is not apparent.  However
as $\hbar\rightarrow 0$ we find at {\em t}
\begin{eqnarray*}
-\frac{\partial S}{\partial t}=H_c (x'_{rt}, p'_{rt}) \hspace{0.5cm}\mbox{and}\hspace{0.5cm} p'_{rt}=\frac{\partial S}{\partial x'_{rt}}.
\end{eqnarray*}
While at $t_0$ we have
\begin{eqnarray*}
\frac{\partial S}{\partial t_0}=H_c (x''_{rt_0}, p''_{rt_0}) \hspace{0.5cm}\mbox{and}\hspace{0.5cm} p''_{rt_0}=-\frac{\partial S}{\partial x''_{rt_0}}.
\end{eqnarray*} 

The significance of Dirac's choice now becomes clearer.  The transition amplitude $\langle x'_t|x''_{t_0}\rangle$ becomes a local momentum in the classical limit.   Recall that in classical mechanics, the equations of motion emerge from infinitesimal contact transformations.  For example, consider the transformation
\begin{eqnarray*}
\{q_j(t), p_j(t)\}\rightarrow \{Q_j(t'), P_j(t')\}.
\end{eqnarray*}
In time $\Delta t$ we have
\begin{eqnarray*}
Q_j-q_j=\Delta q_j=\phi_j(q_j, p_j)\Delta t\quad\mbox{and}\quad P_j-p_j=\Delta p_j=\psi_j(q_j,p_j)\Delta t,
\end{eqnarray*}
so that
\begin{eqnarray*}
Q_j=q_j+\phi_j\Delta t\quad\mbox{and}\quad P_j=p_j+\psi_j\Delta t.
\end{eqnarray*}
A contact transformation is defined to satisfy 
\begin{eqnarray*}
\sum(P_jdQ_j-p_jdq_j)=dS(q_j, p_j)
\end{eqnarray*}
where $S(q_j, p_j)$ is the generating function of the canonical transformation, and as is well known, $S(q_j, p_j)$ is the classical action.  If $dS(q_j, p_j)$ is exact, then
\begin{eqnarray*}
P_j=\frac{\partial S}{\partial Q_j}\quad\mbox{and}\quad p_j=-\frac{\partial S}{\partial q_j}.
\end{eqnarray*}
This immediately recalls the classical Hamilton-Jacobi equation (\ref{eq:CHJ}) where the local momentum is defined as $p=\partial S/\partial x$.  The fact that the real part of the Schr\"{o}dinger equation produces a modified Hamilton-Jacobi equation should immediately raise the question ``What is the significance of the introduction of $e^{iS/\hbar}$?"  

A study of the covering group of the symplectic group 
provides the answer.  The introduction of $e^{iS/\hbar}$ lifts the symplectic structure onto the covering space~\cite{mdg17, mdgbh11}.  We then see that, in some sense, the classical dynamics evolves in the symplectic space while the corresponding quantum phenomena evolve in the covering space.

\section{Quantum Trajectories}

 \subsection{The Problem is Non-commuting Operators} 
 
 With this new structure in mind let us return to the question ``How does a particle get from $A$ to $B$?"  Classically the answer is easy; you can track its position in time knowing its momentum and position simultaneously. The trajectory is then well captured  by the classical canonical formalism.

 In quantum theory we have operators and eigenvalues.  If we try to use the eigenvalues we are halted by the fact that the eigenvalues of the position operator and the eigenvalues of the momentum operator are not simultaneously definable.  Only one becomes definable if we make a measurement.  But we don't want to make a series of measurements between $A$ and $B$ to answer the question.  Mott~\cite{nm29} has 
 already given an answer to this question using the standard view. We are developing a different view.
 
  Dirac~\cite{pd45} began to develop this different view by suggesting we  go to the Heisenberg picture.  Here the operators become $q$-numbers and carry the time dependence.  More importantly, both the position $q$-number and the momentum $q$-number are simultaneously well defined.  If we can give the $q$-numbers an ontological meaning as Aharonov {\em et al.}~\cite{yatlec15} suggest, then it will be possible to discuss a `quantum trajectory'.  Indeed that was what Dirac~\cite{pd45} was already attempting to do when he introduced a method for defining general functions of non-commuting observables in such a way that enables one to discuss trajectories for the motion of a quantum particle.

The method depends on giving an ontological meaning to transition  amplitudes.
 Consider the expression
$\langle X'|f(\hat A\hat B\hat C\hat D)|X\rangle$,
where $| X\rangle$ is the fixed vector and $(\hat A\hat B\hat C\hat D)$ a set of $q$-numbers.  If the $q$-numbers are all mutually commuting, the meaning of the expectation value is clear. Just replace each $q$-number with its eigenvalue and the expression gives the expectation value for the system to have that particular set of eigenvalues in that time order.  

If they do not commute, the eigenvalues cannot be simultaneously specified.  However in the Heisenberg algebra these operators depend on time  and, in the non-relativistic theory, all operators that occur at different times commute\footnote{We will not discuss the relativistic theory in this paper.}.  Thus if we write them in a time order, then a meaning can be given to the expectation value.  

For example suppose we consider a set of points $q_1,q_2\dots q_n$ at different times $t_1 ,t_2, \dots t_n$, we will then find
\begin{eqnarray}
	\langle X'|f(q_1\dots q_n)|X\rangle=\int f(q_1\dots q_n)\langle X'|q_n\rangle dq_n\langle q_n| q_{n-1}\rangle dq_{n-1}\dots dq_{1}\langle q_1| X\rangle.
\end{eqnarray}
Note that we will obtain different expectation values, each depending on the order imposed on the $q$-numbers. 

If $f=1$  for $q_1$ in the ranges $q_1$ to $q_1+dq_1$, $q_2$ in the range $q_2$ to $q_2+dq_2$ etc., and zero otherwise, we get
\begin{equation*}
	\langle X'|f|X\rangle=\langle X'|q_n\rangle dq_n\langle q_n|q_{n-1}\rangle dq_{n-1}\langle q_{n-1}|q_{n-2}\rangle d q_{n-2}\langle q_{n-2}|q_{n-3}\rangle\dots dq_1\langle q_1|X\rangle.
\end{equation*}

To give this a more specific meaning, let us replace  $ |X\rangle $ by $ |Q(t) \rangle$ at the initial time $t$ and $|X'\rangle$ by $ |q(t')\rangle $ at a final time $t'$.  In this way a chain of time-ordered points in space has been defined. Moreover we have the expectation value of an ordered chain of events which is composed of a set of small transition  amplitudes $\langle q_{i+1}|q_{i}\rangle$. If we make the assumption that the Heisenberg algebraic approach describes individual processes, we can interpret $\langle q|Q\rangle$ as giving us the probability amplitude of a zig-zag path connecting the points $Q,q_1, \dots, q_{n-1}, q_n, q$.  This means that we now have a way of discussing the path of a single quantum particle. 

\subsection{Evaluation of the Infinitesimal Propagators $\langle q_{i+1}|q_{i}\rangle$}

Nakahara~\cite{mn90} writes
\begin{eqnarray*}
\langle q_{i+1},t_{i+1}|q_i,t_i\rangle=\langle q_{i+1}|e^{-iH(t_{i+1}-t_i)/\hbar}|q_i\rangle
\end{eqnarray*}
but we will use the more general form assumed by Dirac~\cite{pd45} and Feynman~\cite{rf48}, namely
\begin{eqnarray}
\langle q_{i+1}|q_{i}\rangle=\exp[iS(q_{i+1},q_i)/\hbar] \label{eq:ITA}
\end{eqnarray}
where $S(q_{i+1},q_i)/\hbar$ is some function, real or complex, yet to be identified.  Following Dirac, we have
\begin{eqnarray*}
\langle q|f(\hat q\hat Q)|Q\rangle=f(qQ)\langle q|Q\rangle
\end{eqnarray*}
so that
\begin{eqnarray*}
\langle q|\hat p_r|Q\rangle=-i\hbar\partial_{q_r}\langle q|Q\rangle=\partial_{q_r}S(qQ)\langle q|Q\rangle=\langle q|\partial_{\hat q_r}S(\hat q\hat Q)|Q\rangle.
\end{eqnarray*}
This means we can write
 \begin{eqnarray}
\hat p_r=\partial_{\hat q_r} S(\hat q\hat Q).	\label{eq:pr}
\end{eqnarray}
Similarly
\begin{eqnarray*}
\langle q|\hat P_r|Q\rangle=i\hbar\partial_{Q_r}\langle q|Q\rangle=-\partial_{Q_r}S(qQ)\langle q|Q\rangle=-\langle q|\partial_{\hat Q_r}S(\hat q\hat Q)|Q\rangle 
\end{eqnarray*}
so that 
\begin{eqnarray}
\hat P_r=-\partial_{\hat Q_r} S(\hat q\hat Q).	\label{eq:Pr}
\end{eqnarray}

We now recall a well-known result in the classical Hamilton-Jacobi theory.  For every classical Hamiltonian flow $f_{t't}$ we can associate a generating function $W_{t't}=W_{t't}(q,Q)$, which is a solution of the classical Hamilton-Jacobi equation.  This generating function then defines a pair of canonical momenta so that
\begin{eqnarray*}
(q,p)=f_{t't}(Q,P)\Longleftrightarrow\left \{\begin{array}{c} p=\partial_q W_{t't}(q,Q)\\P=-\partial_{Q} W_{t't}(q,Q).\end{array}\right.
\end{eqnarray*}
Notice here the close relationship between the quantum and classical treatments.   It turns out that $W(qQ)$ is identical to $S(qQ)$.  For example the generating function for the free particle Hamiltonian $H=\bm p^2/2m$ is
\begin{eqnarray*}
W_{t't}(\bm r',\bm r)=m\frac{(\bm r'-\bm r)^2}{2(t'-t)},
\end{eqnarray*}
while the quantum free particle Green's function is
\begin{eqnarray*}
G(\bm r' ,\bm r,t',t)=\left(\frac{m}{2\pi i\hbar(t'-t)}\right)^{3/2}\exp\left(\frac{i}{\hbar}.\frac{m(\bm r'-\bm r)^2}{2(t'-t)}\right).
\end{eqnarray*}
Also, the generating function for the one-dimensional classical harmonic oscillator with $H= (p^2+m^2\omega^2 x^2)/2m$ is
\begin{eqnarray*}
W_{t't}( x', x)=\frac{m\omega}{2\sin\omega(t'-t)}\left((x^{'2}+x^{2})\cos\omega(t'-t)-2x'x\right),
\end{eqnarray*}
while the quantum harmonic oscillator Green's function is
\begin{eqnarray*}
G(x' ,x ,t', t)=\sqrt{\frac{m\omega}{2\pi i\hbar\sin \omega(t'-t)}}\hspace{5cm}\\\times\exp\left[\frac{im\omega}{2\hbar\sin\omega(t'-t)}\left((x^{'2}+x^{2})\cos\omega(t'-t)-2x'x\right)\right].
\end{eqnarray*}
In other words we find that in general in classical mechanics $W_{t't}(x',x)$ generates the Hamiltonian flow $f_{t't}$ where $\left(x'(t'),p'(t')\right)=f_{t't}\left(x(t),p(t)\right)$.  Whereas in quantum mechanics $G_{t't}(x',x)$ generates $U_{t't}$ where $\psi(x', t')=U_{t't}\psi_0(x,t)$.  The relation between the two is 
\begin{eqnarray*}
G_{t't}(x', x)\sim\exp[ iW_{t't}(x',x)/\hbar].
\end{eqnarray*}

\subsection{Particle or Process?}

The fact that the $q$-numbers, the elements of the quantum algebra, are being given precedence over $c$-numbers, means we must also be prepared to change  the notion of a particle from its familiar classical local `rock-like' meaning to a more subtle form of quasi-local, semi-autonomous structure of energy-momentum.  The classical image has already been challenged by special relativity where  there is no concept of rigidity.  An extended particle is envisaged as a world tube of events and processes.  The standard model refines that image further and treats the nucleon in terms  of a small fraction of valence quarks bathed in a sea of quark-antiquark pairs, together with many gluons.  It is a complex `hive' of activity with processes very far removed from any simple classical concepts, so why not try to make sense of the algebraic formalism without pinning it to outdated concepts.

Indeed the mathematical structure  suggested by Dirac was shown by Feynman to satisfy Huygens' principle in a sense described by the Green's function approach.  The algebraic equivalent is an automorphism of the form 
\begin{eqnarray*}
\hat A'=\hat M \hat A\hat M^{-1}.
\end{eqnarray*}
In general mathematical terms we have a two-sided module structure.  In intuitive terms this transformation is more in the nature of a {\em metamorphism} than the point-to-point transformation one uses in classical physics~\cite{db80} . 

The importance of this change in language is hidden in the formal structure of the mathematics.  In the Schr\"{o}dinger picture one is  faced with ray representations and, in consequence, we are faced with projective representations of the symmetry groups.  For example, Haag~\cite{rh92} points out that while classically, the Poincare 
group plays a crucial role, in quantum theory it is the covering group that contains important features like spin that play no role in classical physics.  Indeed it is the covering group of the symplectic group that plays a vital role in the Feynman approach, the significance of which seems to have been missed by the physics community in general. 

The spin structure of the rotation group has, of course, played an essential and major role in physics.  The Pauli spin matrices and the Dirac gamma matrices play a central role in atomic and particle physics.  Furthermore the fact that the spinor changes sign under a $2\pi$ rotation is an indication that we are dealing with a two-fold representation, a representation that has physical consequences that have been experimentally confirmed~\cite{swrc75}.
 Mathematically the easiest way to describe the properties of the spin structure is through the orthogonal Clifford algebra~\cite{pl97}.
This algebra contains the Clifford group in which the rotations of objects on the algebra are given by the formula 
\begin{eqnarray*}
A'=BAB^{-1}
\end{eqnarray*}
where $A$ is some element of the algebra, say, a vector and  $B$ is a product of  a pair of generators. In other words the algebra forms a bimodule where rotations are metamorphisms.

What is not so familiar to physicists working in quantum theory is that the symplectic group, the group from which one can generate Hamilton's equations of motion, has a covering group, the metaplectic group and its non-linear generalisation.  Crumeyrolle~\cite{ac90} has shown that this gives rise to an 
algebraic structure that is analogous to the orthogonal Clifford algebra,  namely the symplectic Clifford algebra.  It is this structure that provides an algebraic method for handling the double cover of the symplectic group.  Furthermore, it is in this structure that the Feynman path integral method takes on a geometric meaning, just as the spin structures take on geometric meaning in the orthogonal Clifford algebra structure~\cite{cdal03}.

\subsection{Geometric Algebras and Feynman's Path Integral Method}

 Feynman defines a propagator through the relation
\begin{eqnarray*}
\psi(x_{k+1},t')=\int_\Omega K(x_k, x_{k+1}, t, t')\psi(x_k,t)dx_k
\end{eqnarray*}
where the integral is taken over a surface $\Omega$ and $K(x_k, x_{k+1}, t, t')$ is defined by
\begin{eqnarray*}
K(x_k, x_{k+1}, t, t')=\exp[i S(x_k, x_{k+1}, t, t')/\hbar].
\end{eqnarray*}
Here $S(x_k, x_{k+1}, t, t')$ is the classical action between the two points $(x_k, t)$ and $(x_{k+1}, t')$.  One way to look at this is that we are summing over the phases of the secondary waves leaving each point on the surface of $\Omega$ and arriving at the point $(x_{k+1}, t')$.  The curious feature of this is that the classical action determines the phases. 

 The clue to the resolution of this mystery lies in the section where we point out that the classical symmetry groups are replaced by  their covering groups.  In classical mechanics the relevant symmetry group is the symplectic group of canonical transformations.  Here the Hamiltonian flows which satisfy the Hamilton-Jacobi equation define an ensemble of classical trajectories.  The generator of the Hamilton-Jacobi equation is the classical action.  What one can show ~\cite{mdg17} is that
  $\exp[iS(x_k, x_{k+1}, t, t')/\hbar]$ is the generator of the covering group of the symplectic group, namely, the metaplectic group.  
  
  Indeed Guillemin and Sternberg~\cite{vgss84} have shown that the Schr\"{o}dinger equation
 appears as a one-parameter sub-group in the covering group.  This means there is a much closer relationship between classical and quantum motion than is usually assumed.  As shown in de Gosson and Hiley~\cite{mdgbh11} the quantum motion is the lift of the classical motion so there is a close relation between the classical and quantum worlds.  In fact there is only
one world.  In other words classical and quantum phenomena are different aspects of the one world.  It was this aspect that Dirac was pointing to in his 1945 paper~\cite{pd45}, the paper from which Feynman~\cite{rf48} obtained his inspiration for the sum-over-paths method.

\subsection{How does a Quantum Particle get from $A$ to $B$?}

With that background let us now return to consider Dirac's proposal that it is possible to discuss how a quantum particle gets from $A$ to $B$ using the non-commutative structure.  Note that this is very much against the standard view expressed in Landau and Lifshitz~\cite{llel13}  that  `in quantum mechanics
 there is no such concept as the path of a particle'.  How then does a quantum particle get from $A$ to $B$? 
 
 Suppose we inject the particle at time $t$ into a volume $\Delta V$ which is large enough to be untroubled by the uncertainty principle.  It is then found some distance away in a volume  $\Delta V'$  at a later time $t'$.  In the standard approach we are restricted to statistical methods, so we can only talk about the probability $|\psi(x',t')|^2$ of finding the particle in $\Delta V'$ at the time $t'$.  Clearly we only have the probability current $\bm j=\hbar[\psi^*(\bm \nabla\psi)-(\bm\nabla\psi^*)\psi]/2mi$ at our disposal to account for how the particle ends up in volume $\Delta V'$.  
 
 However this does not enable us to say anything about how a single particle gets from $\Delta V$ to $\Delta V'$.  
 To avoid problems with the uncertainty principle, consider a small volume $\Delta V$ surrounding the point $q$.  Imagine a sequence of particles emanating from a point in  $\Delta V$, each with a different momentum, so that over time we have a spray of all possible momenta emerging from the volume $\Delta V$.  Similarly there is a spray of momenta arriving at the small volume $\Delta V'$ surrounding the point $q'$.  
  
\begin{figure}[h] 
     \centering
     \includegraphics[width=4in]{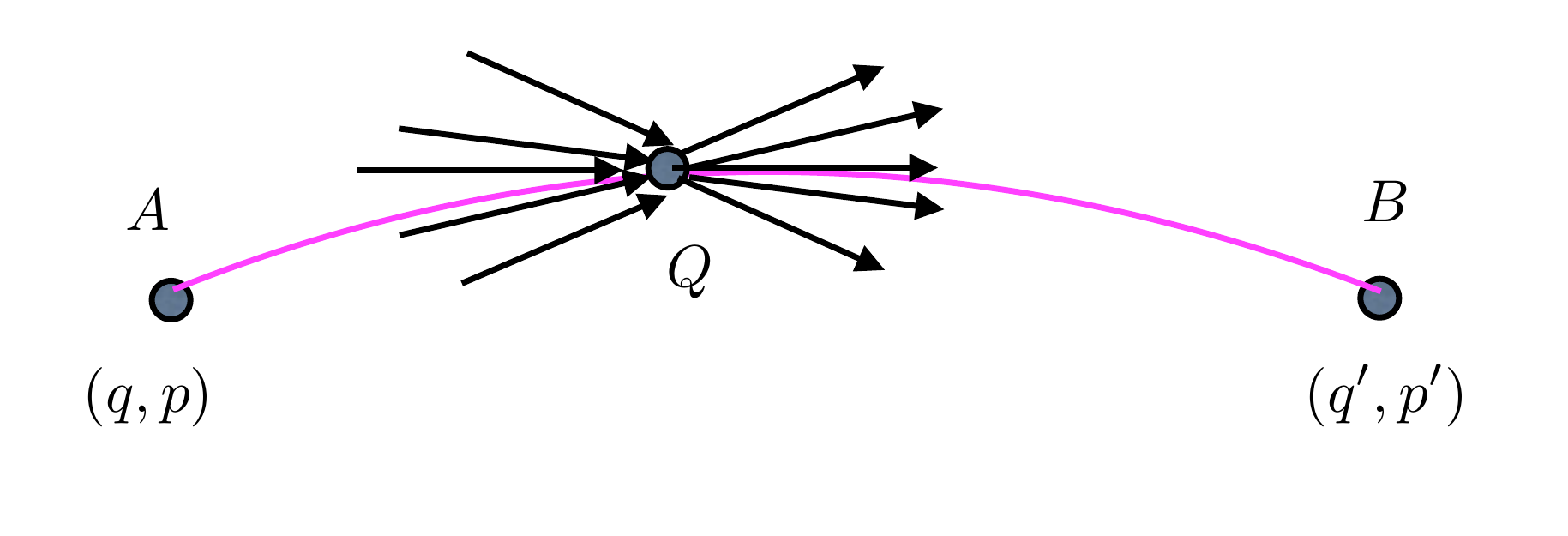} 
     \caption{Behaviour of the momenta sprays at the midpoint of $\langle q', t'| q, t\rangle_\epsilon$}
     \label{fig:spray}
  \end{figure}

  Better still, let us consider a small volume surrounding the midpoint $Q$.  At this point there is a spray of momenta arriving and a spray leaving a volume $\Delta V(Q)$ as shown in Figure \ref{fig:spray}.
  To see how the local momenta behave at the midpoint $Q$, recall that for small time differences $t'-t=\epsilon$, we have for the propagator of a free particle,
  \begin{eqnarray}
  S_\epsilon(q',q)=\frac{m}{2}\frac{(q'-q)^2}{\epsilon}	\label{eq:act}
  \end{eqnarray}
  which is obtained from the classical Lagrangian.  Then we have the  momentum TA
  \begin{eqnarray}
  P_Q(q',q)=\frac{\partial S_\epsilon(q', q)}{\partial Q}=\frac{\partial S_\epsilon(Q,q)}{\partial Q}+\frac{\partial S_\epsilon(q', Q)}{\partial Q}.	\label{eq:momXa}
  \end{eqnarray}
  Using  (\ref{eq:act}), we find
  \begin{equation}
  P_Q(q',q)=m\left[\frac{(q'-Q)}{\epsilon}-\frac{(Q-q)}{\epsilon}\right]=p'_Q(q',q)+p_Q(q',q).  \label{eq:momX}
  \end{equation}
  The RHS of this equation comprises exactly the relations that Dirac~\cite{pd45} obtained in equations (\ref{eq:pr}) and  (\ref{eq:Pr}) above.  Not surprisingly, it is also
   exactly the momentum TA that Feynman~\cite{rf48} obtains in his equation (48)  at the point $Q$ which lies between the two neighbouring points separated in time by $\Delta t=\epsilon$. 
   
    Notice that in the limit of $\epsilon\rightarrow 0$, equation (\ref{eq:momX}) comprises two `derivatives' at $Q$, namely
  \begin{eqnarray*}
  D_Q(\mbox{Backward})=\lim_{\epsilon\rightarrow 0}\frac{(Q-q)}{\epsilon}\hspace{1cm}  D_Q(\mbox{Forward})=\lim_{\epsilon\rightarrow 0}\frac{(q'-Q)}{\epsilon}.
  \end{eqnarray*}
  Such derivatives are associated with a general stochastic process where the `trajectory' joining the two points $q$ to $q'$ is continuous, but the derivatives are not.   This situation is known to arise in Brownian motion~\cite{nw66}.  Indeed these very derivatives were used by Nelson~\cite{en66}  in his derivation of the Schr\"{o}dinger equation from an underlying stochastic process. (See also the discussion in Bohm and Hiley~\cite{dbbh89} and  Prugove\u{c}ki~\cite{ep82} for alternative views.)

The meaning of the non-continuous derivatives here is clear;
the basic underlying quantum process connecting infinitesimally neighbouring points is an {\em intrinsically} random process, but at this stage the precise form of this stochastic process is unclear.  However the spray of possible momenta emanating from a region cannot be completely random since, as Feynman has shown, the transition amplitudes satisfy the Schr\"{o}dinger equation under certain assumptions.  Some clues as to the precise nature of this distribution have already been supplied by Takabayasi~\cite{tt54} and Moyal~\cite{jm49}, clues which we will now exploit.

We are interested in finding  the average behaviour of the momentum, $P_Q$, at the point $Q$.  This means we must determine the spray of momenta that is consistent with the wave function $\psi(Q)$ at $Q$.  But we have two contributions, one coming from the point $q$ and one leaving for the point $q'$.   Feynman's proposal~\cite{rf48} that we can think of $\psi(Q)$ as `information coming from the past' and $\psi^*(Q)$ as `information coming from the future', will be used here as this suggests that we can write
\begin{eqnarray*}
\lim_{q\rightarrow Q}\psi(q)=\int\phi(p)e^{ipQ} dp\hspace{0.5cm}\mbox{and}\hspace{0.5cm} \lim_{Q\rightarrow q'}\psi^*(q')=\int \phi^*(p')e^{-ip'Q}dp'.
\end{eqnarray*}
The $\phi(p)$ contains information regarding the probability distribution of the incoming momentum spray, while $\phi^*(p')$ contains information about the  probability distribution of the outgoing momentum spray.  These wave functions must be such that in the limit $\epsilon\rightarrow 0$ they are consistent with the wave function $\psi(Q)$.
Thus we can define the mean momentum, $\overline {P(Q)}$, at the point $Q$ as
\begin{eqnarray}
\rho(Q)\overline {P(Q)}=\int\int P\phi^*(p')e^{-ip'Q}\phi(p)e^{ipQ}\delta(P-(p'+p)/2)dPdpdp'		\label{eq:MMPP}
\end{eqnarray}
where $\rho(Q)$ is the probability density at $Q$.   We have added the restriction $\delta(P-(p'+p)/2)$ because we are using the diffeomorphism $(p,p')\rightarrow [(p'+p)/2,(p'-p)]$.  It is immediately seen that equation (\ref{eq:MMPP}) can be put in the form   
 \begin{eqnarray}
    \rho(Q)\overline {P(Q)}=\left(\frac{1}{2i}\right)[(\partial_{q_1}-\partial_{q_2})\psi(q_1)\psi(q_2)]_{q_1=q_2=Q},	\label{eq:MMXX}
  \end{eqnarray}
  a form that appears in Moyal~\cite{jm49}.
  
  If we write the wave function in polar form, we find that  $\overline {P(Q)}$ is just the local momentum $P_B=\nabla S$ that appears in the Bohm interpretation.  Since $P_B$ is used to calculate the Bohm trajectories, there must be a close relationship between these trajectories and Feynman paths.  If we assume each evolving quantum process, which we will call a particle, actually follows a Feynman stochastic path then a Bohm trajectory can be regarded as an ensemble average of many such paths. Notice however, this gives a very different picture of the Bohm momentum  from the usual one used in Bohmian mechanics~\cite{ddst09}.  It is not the
momentum of a single `particle' passing the point $Q$, but the mean {\em momentum flow} at the point in question.  

The question remains as to the nature of the underlying reality.  Is it particulate in nature or is it a more subtle notion of a quantised process involving a novel organisation of energy and momentum?  The Bohm approach was taken as support for a particle-like picture, even though the appearance of the quantum potential suggested that there was an element of non-locality present, an element of the wholeness Bohr talked about.

The identification of the canonical relation $P_B=\nabla S$, an unjustified assumption in Bohm's original paper~\cite{db52}, was always a worrying feature of the approach.  Now we see that it has its origins
in the averaging over a deeper fundamental non-commutative stochastic process, being related to the infinitesimal  transition amplitude shown in equation (\ref{eq:ITA}).

Further support for this view comes from field theory itself.  If we  treat the Schr\"{o}dinger $\psi(x,t)$ as a field described by a Lagrangian
\begin{eqnarray}
{\cal L}== - \frac{1}{2m}\nabla \psi^{*}\cdot\nabla \psi+\frac{i}{2}[(\partial_{t}\psi)\psi^{*}-(\partial_{t} \psi^{*})\psi]-V\psi^{*}\psi,	\label{eq:Lagrangian}
\end{eqnarray}
then the energy-momentum tensor can be written as
\begin{eqnarray}
T^{\mu\nu}=- \left\{\frac{\partial {\cal{L}}}{\partial(\partial^{\mu}\psi)}\partial^{\nu}\psi+\frac{\partial {\cal{L}}}{\partial(\partial^{\mu}\psi^{*})}\partial^{\nu}\psi^{*}\right\}-{\cal L}\delta^{\mu\nu}.	\label{eq:EM}
\end{eqnarray}
From this we find 
\begin{eqnarray*}
T^{0j}=\frac{i}{2}\left[\psi^{*}\partial^{j}\psi - \psi\partial^{j}\psi^{*}\right]
\end{eqnarray*}
with $\partial^{j}=-\nabla$.  This is immediately seen to give $P_B=\nabla S$.  Thus the Bohm momentum, and hence the Bohm energy, is the field energy-momentum.  In this way we see that the Bohm approach makes the time evolution of a quantum process {\em look like} a particle following a trajectory.  Schwinger~\cite{js51} had already shown, in a different way, how from relativistic field theory one
can be led to a set of dynamical equations which look like a ``particle" evolving in a proper time coordinate. We will show how to extend this idea to the fully algebraic theory in a later publication.

\section{Conclusion}

Replacing the variables of classical mechanics by $q$-numbers as suggested by Dirac~\cite{pd45} enables us to propose a radical ontology for the underlying quantum processes.  Moreover this contains  classical mechanics as a natural limit without the need for decoherence.  Bohmian mechanics becomes an intermediary showing that the so-called `particle trajectories' are an ensemble average of stochastic Feynman paths, where these paths describe a natural stochastic evolution of a process involving quanta of energy and momenta.  The exact nature of this process is still unclear but we believe the algebraic method opens up new ways of exploring this underlying structure.

Already quantum loop gravity~\cite{cr01} provides a radical new view of this structure.  Spin nets and their generalisation open up the possibility of a `quantum space-time' where space-time is not taken as an {\em a priori} given, but rather the properties of space-time emerge from this process.  This approach then re-visits the original ideas proposed by Dirac~\cite{pd45} and Feynman~\cite{rf:process}, that were later abandoned in favour of an algorithm plagued with infinities.  Yes, the algorithm was used very successfully for quantum electrodynamics and high energy physics, but it fails completely for quantum gravity.  The new possibilities not only open up a new approach to quantum gravity, but also throw new light on the old interpretational problems and in such a way that they begin to fade away.

In the algebraic approach all the information contained in the wave function is seen to be encoded in the algebra itself, in the form of the elements of the left ideals.  Representing these elements by vectors in an abstract Hilbert space certainly simplifies the mathematics as an algorithm but leaves us with the likes of schizophrenic cats and the century-old problem of the collapse of the wave function.  We now propose that there are actual individual processes and that these are basically stochastic.  By that we mean that  Newton's first law does not hold for the individual quantum processes,  nor  does it require a sub-quantum medium as originally proposed by de Broglie~\cite{ldb64}.  Rather, the stochasticity is of such a nature that Newton's first law emerges at the classical level. 

This leaves open the question as to the detailed nature of this underlying process.  Will the ideas underlying quantum loop gravity provide the answers?  Or will some more radical approach involving a fractal space-time, a notion proposed by Nottale~\cite{ln94} be required?  These questions will be taken up in a later paper.

 \vspace{0.3cm}
\noindent {\bf Acknowledgements}

One of us (BJH) would like to thank the Fetzer Franklin  Fund  for their financial support.



\bibliography{myfile}{}
\bibliographystyle{plain}

\end{document}